\documentclass[aps,prc,a4paper,twocolumn,showpacs,showkeys,floatfix, superscriptaddress]{revtex4-1}

\usepackage[dvips]{epsfig}
\usepackage{color}
\usepackage{amsmath}
\usepackage{multirow}

\newcommand{\figwidth}{\columnwidth}

\newcommand{\vect}[1]{\mathbf{#1}}
\newcommand{\cuen}{Cu(en)(H$_2$O)$_2$SO$_4$}
\begin{document}

\title{Splitting of antiferromagnetic resonance modes  in the quasi-two-dimensional collinear antiferromagnet Cu(en)(H$_2$O)$_2$SO$_4$}

\author{V. N. Glazkov}
\email{glazkov@kapitza.ras.ru}
\affiliation{ P.L. Kapitza Institute for Physical Problems, RAS, Kosygina 2, Moscow 119334, Russia}
\affiliation{International Laboratory for Condensed Matter Physics, National research university ``Higher School of Economics'', Myasnitskaya str. 20, 101000 Moscow, Russia}

\author{Yu. V. Krasnikova}
\affiliation{ P.L. Kapitza Institute for Physical Problems, RAS, Kosygina 2, Moscow 119334, Russia}
\affiliation{International Laboratory for Condensed Matter Physics, National research university ``Higher School of Economics'', Myasnitskaya str. 20, 101000 Moscow, Russia}

\author{I. K. Rodygina}
\affiliation{ P.L. Kapitza Institute for Physical Problems, RAS, Kosygina 2, Moscow 119334, Russia}
\affiliation{Faculty of Physics, National research university ``Higher School of Economics'', Myasnitskaya str. 20, 101000 Moscow, Russia}

\author{J. Chovan}
\affiliation{IT4Innovations National Supercomputing Center, VSB-Technical University of Ostrava, 17. listopadu 2172/15, CZ 708 33 Ostrava, Czech Republic }
\affiliation{ International Clinical Research Center, St. Anne's University Hospital, Pekarska 53, 656 91 Brno, Czech Republic}

\author{R. Tarasenko}
\affiliation{Institute of Physics, P. J. \v{S}af\'{a}rik University, Park Angelinum 9, 040 00 Ko\v{s}ice, Slovakia}

\author{A. Orend\'a\v{c}ov\'a}
\affiliation{Institute of Physics, P. J. \v{S}af\'{a}rik University, Park Angelinum 9, 040 00 Ko\v{s}ice, Slovakia}

\begin{abstract}
Low-temperature magnetic resonance study of the quasi-two-dimensional
antiferromagnet \cuen{} (en = C$_2$H$_8$N$_2$) was performed down to 0.45~K. This compound orders antiferromagnetically at 0.9~K. The analysis of
the resonance data within the hydrodynamic approach allowed to identify
anisotropy axes and to estimate the anisotropy
parameters for the antiferromagnetic phase. Dipolar spin-spin coupling turns out to be the main contribution to the anisotropy of the antiferromagnetic phase. The splitting of the resonance
modes and its non-monotonous dependency on the applied frequency was
observed below 0.6~K in all three field orientations. Several models were
discussed to explain the origin of the nontrivial splitting and the
existence of inequivalent magnetic subsystems in \cuen{} was
chosen as the most probable source.
 \end{abstract}

\date{\today}



\maketitle

\section{Introduction}

Low-dimensional antiferromagnets are one of the focus topics of modern magnetism. Low dimensionality of the spin system enhances the role of thermal and quantum fluctuations, retarding  magnetic ordering in these systems to the lower temperatures $T_N\ll\Theta$ (here $\Theta$ is a Curie-Weiss temperature) or even fully suppressing it. This yields the extended temperature range of the spin-liquid behavior where short-range spin-spin correlations determine the dynamics of the disordered spin system. ``Freezing'' of this spin-liquid under the effect of weak coupling between the low-dimensional subsystems, additional further-neighbor or anisotropic interactions, external field or applied pressure is of interest since the competing weaker interactions sometimes give rise to  complex magnetic phase diagrams or to the appearance of the unusual (e.g., spin-nematic) phases  \cite{dejong,zapf,fortune,zhitomirskii,starykh}.

Two-dimensional (2D) antiferromagnets are also of interest due to the occurrence of topological excitations induced by the magnetic field and/or the easy-plane spin anisotropy  \cite{bkt1,bkt2}. A crossover between the low- and high-temperature regimes of the spin dynamics appears in the vicinity of  the topological Berezinskii-Kosterlitz-Thouless (BKT) transition accompanied with the formation of the bound pairs of  vortex-antivortex excitations  \cite{bkt3}.

Recently studied antiferromagnet \cuen{} (here (en)$=$C$_2$H$_8$N$_2$) is an example quasi-2D system. Combination of the thermodynamic measurements  \cite{lederova2017,kajnakova2005} and first-principle calculations  \cite{firstprinciples} proved that its spin subsystem can be envisioned as a 3D array of coupled  zig-zag square lattices with the strongest in-plane exchange coupling constant $J/k_B \simeq3.5$~K and the interplane coupling $J'<0.03 J$ \cite{lederova2017}. \cuen{} orders antiferromagnetically at $T_N=(0.91\pm 0.02)$~K, the ordering is accompanied by a sharp $\lambda$-like anomaly in the specific heat and by the appearance of the strong anisotropy in the magnetic susceptibility. The $(H-T)$ phase diagram was discussed in the context of a field-induced BKT transition  \cite{lederova2017,phases-jmmm}. Observed enhancement of the field-induced transition temperatures qualitatively follows predictions for the field-induced BKT transition \cite{bkt1}, this feature is characteristic for quasi-2D antiferromagnets \cite{sengupta}.

In this paper we report the results of the low-temperature electron spin resonance study of the magnetic ordering in \cuen{} down to 0.45~K. Magnetic resonance spectroscopy of the ordered phase probes the $q=0$ magnon spectrum with high energy resolution (routine resolution of 1~GHz corresponds to 0.005~meV) thus giving insight into the structure of the  magnetic phase, type of magnetic ordering, magnetic phase transitions etc. Our observations confirmed collinear ordering in \cuen{} and allowed to unambiguously identify the anisotropy axes and to determine the anisotropy parameters of the ordered phase. Observed anisotropy can be successfully described by dipole-dipole interaction.  We also observed  splitting of the resonance lines in the ordered phase indicating the presence of inequivalent antiferromagnetic subsystems below N\'{e}el temperature.

\section{Experimental details, samples and crystal structure}
\begin{figure}
\epsfig{clip=, width=\figwidth, file=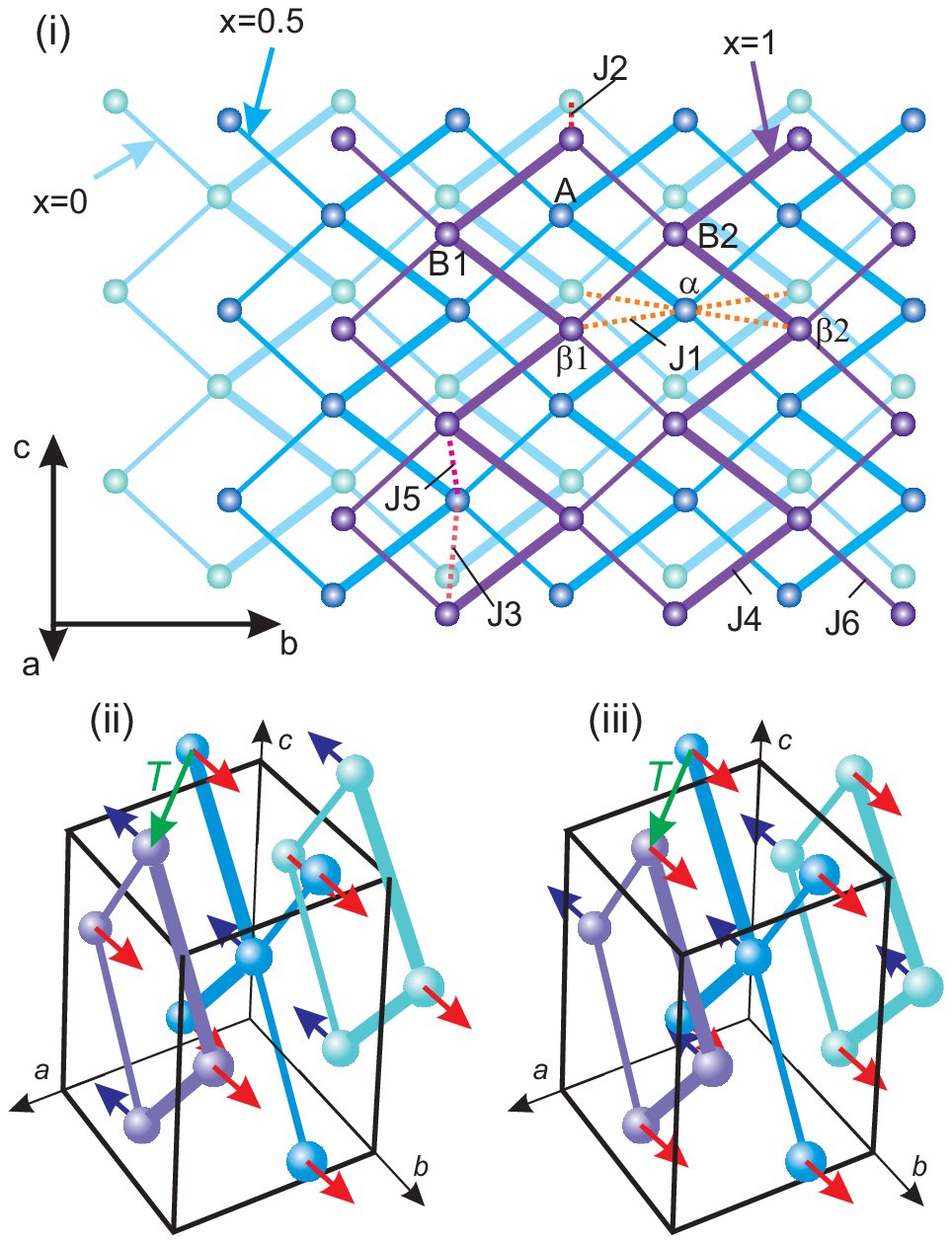}
\caption{(i) Fragment of \cuen{} crystal structure projected on the $(bc)$-plane. Only copper ions are shown, thicker lines correspond to the strongest in-plane couplings. Layers shown by different shade of color are formed by atoms with fractional $x$ coordinate 0, 0.5 and 1.0 correspondingly. Relevant in-plane and inter-plane couplings are shown according to Ref.~\cite{firstprinciples}. Ions within the adjacent $(ab)$ planes coupled by the DM interaction are marked as ``A'', ``B1'', ``B2'' (``$\alpha$'', ``$\beta1$'', ``$\beta2$'').  (ii), (iii) Schemes of antiferromagnetic (ii) and ferromagnetic (iii) stacking of the antiferromagnetically ordered $(bc)$-planes, blue and red arrows show ordered magnetic moment orientation, green arrows show elementary translation $\vect{T}=(\vect{a}+\vect{b})/2$ linking the neighboring planes.}
\label{fig:planes}
\end{figure}

Electron spin resonance (ESR) experiments were performed using a set of home-made transmission type spectrometers covering  frequency range from 4 to 120~GHz, some of the spectrometers were equipped with ${}^3$He-vapor pumping cryostats allowing to reach temperature as low as 0.45~K. Magnetic fields up to 12 T were created by a compact superconducting cryomagnets. Resonance absorption was recorded as a dependency of the transmitted microwave power on the slowly swept magnetic field.

For the most of our experiments samples were mounted on the bottom of the cylindrical (above 20~GHz) or rectangular (9-20~GHz) multimode microwave cavity. A small sapphire block was used as a heat link for the sample orientations preventing the plane-on-plane sample mounting. Low-frequency ESR experiments at the frequencies 4-8~GHz were performed in $\vect{H}||b$ orientation only with the help of quasi-toroidal resonator.

\cuen{} (abbreviated CUEN for short) crystals were grown by the same technique as the samples used in Ref.~\cite{lederova2017}. \cuen{} crystallizes in the base-centered monoclinic space group $C^6_{2h}$. The $(bc)$-planes are stacked in $(1/2,1/2,0)$ direction. Primitive unit cell contains two copper ions, positions of these ions are linked by inversion. Second order rotational axis is parallel to the $b$ axis and passes through the copper ions. Fragment of the crystal structure of CUEN is shown in Figure~\ref{fig:planes}. As-grown crystals of \cuen{} are blue-colored elongated thin plates with the long edge parallel to the $a$ direction and the sample plane normal to the $b$ direction. Samples shape allowed easy positioning of the sample at $\vect{H}||a,b,c^*$.

The formation of 2D planes of exchange coupled spins was confirmed by the characteristic behavior of the specific heat and magnetization and by first principles calculations  \cite{lederova2017,kajnakova2005,firstprinciples}. Relevant exchange bonds are shown in Figure \ref{fig:planes}. The in-plane coupling values are (in the notations of Figure \ref{fig:planes} and Ref.~\cite{firstprinciples}) $J_4=3.4...3.7$~K and $J_6=0.35 J_4$ \cite{lederova2017}.
Inversion centers in the middle of in-plane copper-copper bonds forbid the in-plane Dzyaloshinskii-Moriya (DM) interaction. However, the inter-plane DM coupling is possible along the $J_1$ and $J_2$ bonds \cite{lederova2017,tarasenko2013}. The same inversion symmetry insures the same orientation of $g$-tensor axes for two copper ions within the primitive unit cell.

\section{Experimental results}
\begin{figure}
\epsfig{clip=, width=\figwidth,file=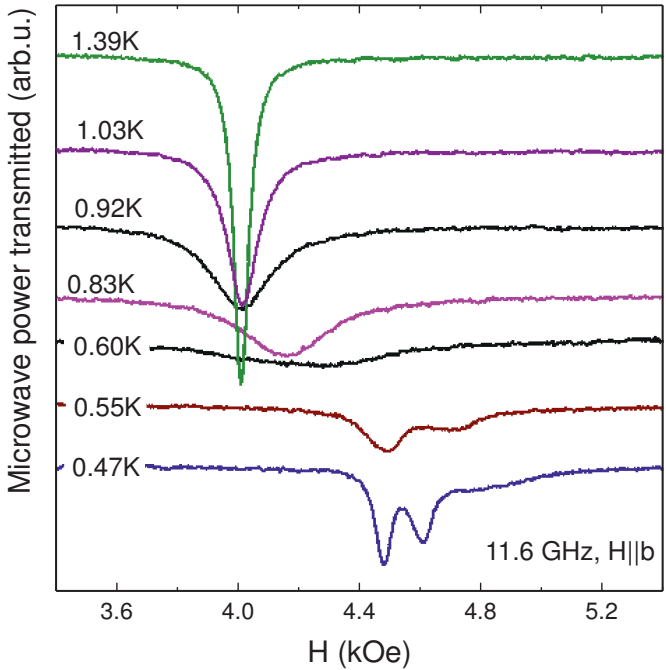}
\caption{Temperature evolution of the resonance absorption in \cuen{} at low temperatures. $f=11.6$~GHz, $\vect{H}||b$.}
\label{fig:scans}
\end{figure}

\begin{figure}
\epsfig{clip=, width=\figwidth,file=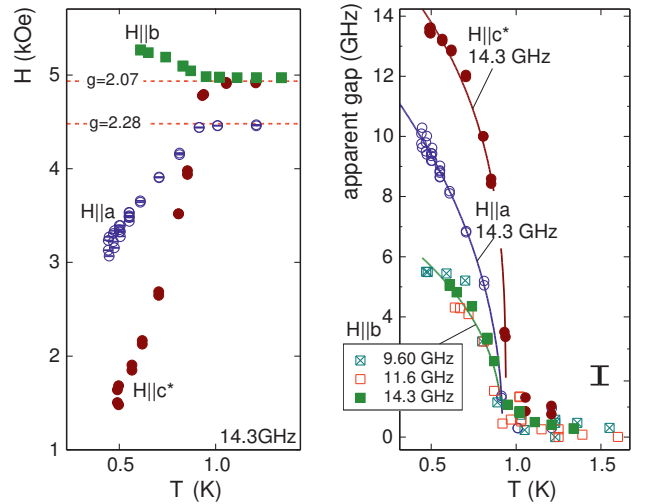}
\caption{(Left) Temperature dependencies of the resonance field for three principal field directions, $f=14.3$~GHz. Horizontal dashed lines mark paramagnetic resonance field for the  $g$-factor values for corresponding field directions. (Right) Temperature dependencies of the apparent gap for three principal field directions. Symbols --- experimental data, solid lines --- phenomenological fits of the temperature behavior of the gap below the transition point as described in the text. The vertical bar shows the error estimate.}
\label{fig:gaps}
\end{figure}

\begin{figure}
\epsfig{clip=, width=\figwidth,file=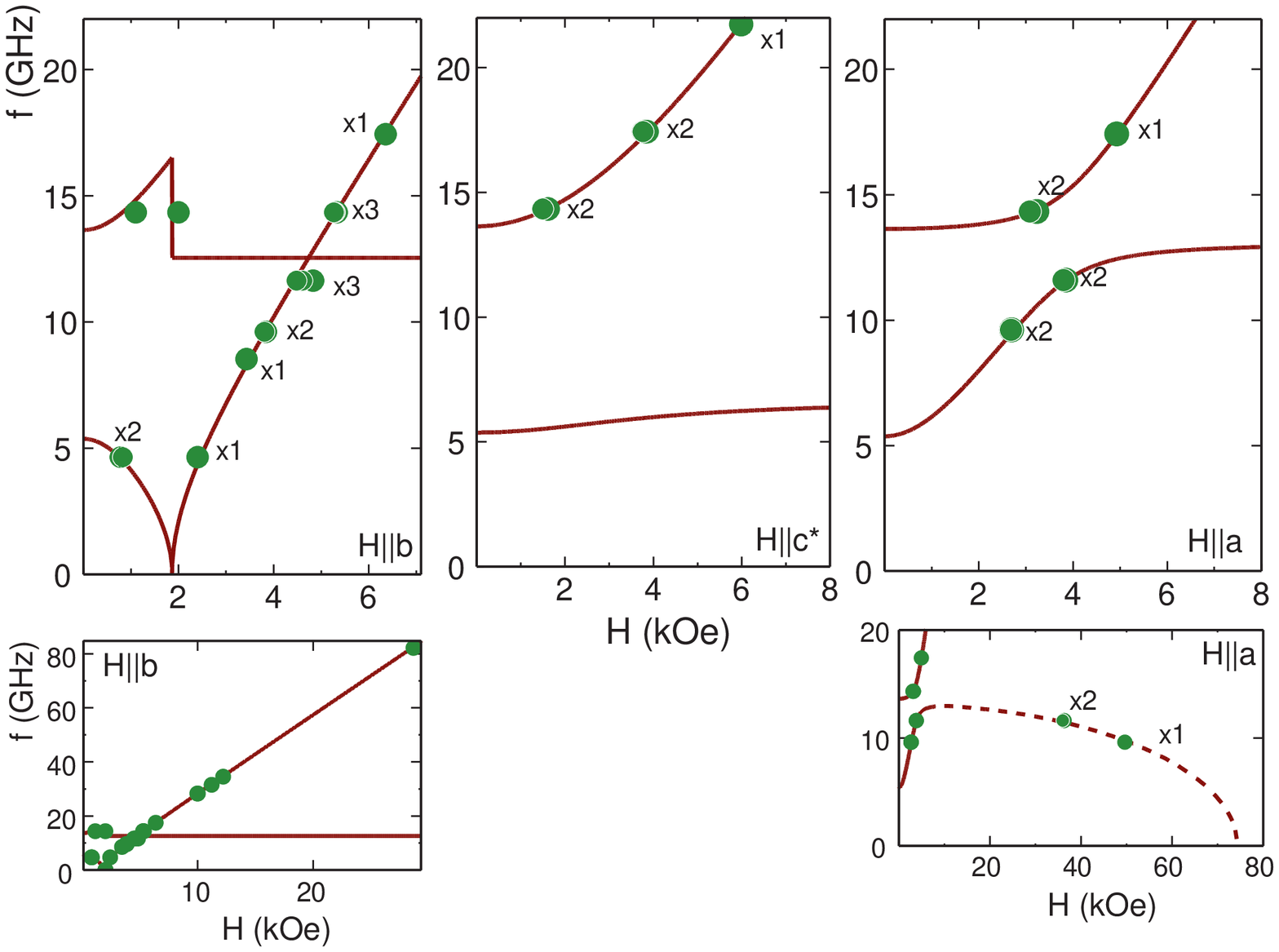}
\caption{Frequency-field diagrams of resonance absorption at the base temperature $T=0.45$~K. Upper row shows low-frequency and low-field parts of $f(H)$, lower row shows higher frequency or higher fields $f(H)$ diagrams for principal field directions. Symbols: experimental data, solid lines: low-field theory as described in the text, dashed line: high-field fit as described in the text. Numbers at the data point 'x1', 'x2', 'x3' show number of split components observed below 0.6~K at a particular frequency.}
\label{fig:f(h)}
\end{figure}

Above the N\'{e}el point we observed  a single-component paramagnetic resonance line with the $g$-factor values determined from 10-120~GHz measurements as $g_a=2.28\pm0.02$, $g_b=2.06\pm0.02$ and $g_{c^*}=2.07\pm0.02$. Found $g$-factor values are in agreement with the earlier results  \cite{kajnakova2005,tarasenko2013}. No splitting of the ESR line was observed at $T>T_N$ both in all principal field orientations and in the control experiment with rotation of the applied field in the  $(ac^*)$-plane performed at the microwave frequency of 72.7~GHz (with the resonance fields around 24~kOe).

Antiferromagnetic transition point is marked by the shift of the resonance absorption from the paramagnetic position (see Figure~\ref{fig:scans}).  Direction of the shift depends on the orientation of the applied field (Figure \ref{fig:gaps}) indicating presence of anisotropy in the ordered phase. For the simple easy axis antiferromagnet  \cite{kubo,goorevich}  the resonance frequency is  $f=\sqrt{(\gamma H)^2\pm\Delta^2}$ (here $\gamma$ is a gyromagnetic ratio and $\Delta$ is a gap in the magnon spectrum) for the field applied perpendicular to the easy axis and along the easy axis above the spin-flop field, correspondingly. Thus, observed increase of
the resonance field for $\vect{H}||b$ confirms earlier identification
of this axis as the easy axis of anisotropy \cite{lederova2017}.

To compare resonance field shift $H(T)$ for different field orientations and different microwave frequencies we have calculated the apparent gap:

\begin{equation}\label{eqn:apgap}
\Delta(T)=\sqrt{|f^2-(\gamma H)^2|}=\gamma \sqrt{|H_{pm}^2-H^2|},
\end{equation}

\noindent here $H_{pm}=f/\gamma$ is the paramagnetic resonance field above the N\'{e}el temperature. The shift of the resonance field below the transition temperature is smooth (see Figure~\ref{fig:gaps}), which indicates continuous development of the order parameter and confirms the second order phase transition. The apparent gap temperature dependency differs for $\vect{H}||a$ and $\vect{H}||c^*$, which indicates that anisotropy is biaxial. Transition temperature determined from the ESR experiment is $T_N=(0.92\pm0.02)$~K, it is in agrement with the known results.

For $\vect{H}||a,c^{*}$ (perpendicular to the easy axis) the apparent gap is proportional to the order parameter \cite{goorevich}. To quantify determined order parameter temperature dependency we fitted $\Delta(T)$ data by the phenomenological law $\Delta\propto (1-T/T_N)^\beta$ in the full temperature range from the base temperature of 0.45~K to $T_N$, the phenomenological exponent values are $0.40\pm 0.02$ and $0.31\pm0.03$ for $\vect{H}||a,c^{*}$, respectively.  The obtained  exponent values  are close  to the critical exponent values for 3D Ising model $\beta_{Ising}^{(3D)}\approx0.327$ \cite{kolesik}, 3D XY-model $\beta_{XY}^{(3D)}=0.3485$ \cite{campostrini} and 3D Heisenberg model $\beta\approx 0.36$ \cite{heis-exp}. All three models could be relevant for CUEN at various fields and temperatures:  magnetization studies at 0.5 K  \cite{lederova2017} revealed the effect of intrinsic spin anisotropies for all field orientations at $H<2$kOe, which corresponds to the Ising model, at higher fields the potential prevalence of the field-induced easy-plane anisotropy can be expected  resulting in the preference of the XY model at lowest temperatures, while at some higher temperatures, a crossover to isotropic Heisenberg behavior occurs \cite{phases-jmmm}.  Similar interpretation of the $\Delta(T)$ dependency  at $\vect{H}||b$ is not possible: within the mean field model  the  position of the resonance absorption for the field applied along the easy axis depends not only on the magnon gap, but also on the temperature-dependent longitudinal susceptibility \cite{kubo}, which has unusual temperature dependency in CUEN \cite{lederova2017}. However, our data can be reasonably fitted by the phenomenological law with  $\beta=0.50\pm0.07$.

We have collected the resonance absorption curves at the base temperature of 0.45~K for different frequencies for $\vect{H}||a,b,c^*$. The final frequency-field diagrams are shown in Figure~\ref{fig:f(h)}. The observed $f(H)$ dependencies are in a qualitative agreement with the known case of a two-sublattice  antiferromagnet with two axes of anisotropy  \cite{kubo}.

Besides of the monotonous shift of the resonance absorption below $T_N$, development of a certain ``fine structure'' of the resonance line was observed on cooling below approximately 0.6~K (Figure \ref{fig:scans}). This splitting of the resonance line was observed in all three field orientations studied, its magnitude is up to 100...150 Oe and is much smaller than the resonance field value. Possible origin of this splitting will be discussed in the following sections, note, however, that the standard model of antiferromagnetic resonance in a collinear antiferromagnet does not allow such a splitting.

\section{Discussion}
\subsection{Frequency-field diagrams analysis \label{sec:f(H)}}

The magnetic resonance spectroscopy is a sensitive and informative method to study antiferromagnetic ordering. At the transition point the single-mode paramagnetic resonance absorption spectrum  transforms into a multi-mode antiferromagnetic resonance (AFMR) absorption spectrum with nonlinear $f(H)$ dependency. For a collinear antiferromagnet  there are always only two low-energy modes  \cite{kubo,goorevich,andmar}, while for a noncollinear antiferromagnet there should be three low energy modes  \cite{andmar} with completely different $f(H)$ diagrams (see, e.g., \cite{prozorova-garnet,zaliznyak,svistov-farutin,tanaka,sosin1,sosin2,glazkov-garnet,glazkov-noncol}). We observe (save for the small splitting of resonance lines which will be addressed later) only two modes of antiferromagnetic resonance (Figure \ref{fig:f(h)}) with $f(H)$ dependencies  typical for a collinear antiferromagnet.  Thus, we can definitely conclude that the antiferromagnetic ordering in \cuen{} is collinear.

Softening of one of the resonance modes in $\vect{H}||b$ at approximately 2~kOe marks a spin-flop transition observed earlier in a low temperature magnetization study  \cite{lederova2017}, this observation proves that the $b$ axis is the easy axis of anisotropy.

Distinct $f(H)$ and $\Delta(T)$ dependencies for $\vect{H}||a,c^*$ (Figures \ref{fig:gaps},\ref{fig:f(h)}) indicate that the anisotropy is biaxial.   The $f(H)$ dependencies for the biaxial collinear antiferromagnet (see, e.g., Ref.~\cite{kubo}) are characterized by two zero-field gaps $\Delta_1>\Delta_2$. Identification of the anisotropy axes from the AFMR data is model-independent: For the field applied  exactly along the hard or second-easy axis one of the eigenmodes is field-independent. The field-dependent AFMR mode corresponds to the oscillations of the order parameter from its equilibrium position along the easy axis toward the field direction and back \cite{kubo,goorevich}. Since the energy cost is smaller for the deviations toward the second-easy axis, the field-dependent mode starts from the lower gap $\Delta_2$ for the field applied along the second-easy axis (close to $\vect{H}||a$ in our case). Consequently, the hard axis of anisotropy (the less favorable orientation of the order parameter ) is close to the $c^*$ axis.

Quantitative analysis of the AFMR $f(H)$ curves was performed within the hydrodynamic approach framework \cite{andmar}. This approach is valid well below the saturation field, this condition is fulfilled for most of our data ($H_{sat} \approx 70$~kOe for CUEN). Low-energy spin dynamics of the collinear antiferromagnet at $T=0$ is described as the uniform oscillations of  the order parameter vector field with the Lagrangian density:

\begin{equation}
\label{eqn:lagrangian}
{\cal L}=\frac{\chi_{\perp}}{2\gamma^2} \left(\dot{\vect{l}}+\gamma \left[\vect{l}\times\vect{H} \right] \right)^2-U_A
\end{equation}

\noindent here unit vector $\vect{l}$ is the collinear AFM order parameter (which is  the normed vector of the staggered magnetization for the sublattices model), $\gamma$ is the gyromagnetic ratio, $\chi_\perp$ is the transverse susceptibility and $U_A(\vect{l})$  is the anisotropy energy depending on the order parameter orientation:

\begin{equation}
U_A=\frac{a_1}{2}l_X^2+\frac{a_2}{2}l_Y^2+\xi \chi_{\perp} (\vect{l}\vect{H}) l_a H_a,
\label{eqn:UA}
\end{equation}

\noindent here first two terms describe conventional biaxial anisotropy, $a_1>a_2>0$,  $X$ and $Y$ being the directions of the hard ($X$) and of the second-easy ($Y$) axes, and the last term describes  axial $g$-factor anisotropy (within the mean-field approach $\xi=\Delta g/g$ \cite{glazkov-bacusio}) with the $g$-tensor principal axis coinciding with the $a$ axis \cite{tarasenko2013}. Due to the low symmetry of \cuen{} only one axis (easy axis $Z$) is pinned to the only second order crystallographic axis $b$ while the orientation of the hard and second easy axes in the $(ac)$-plane is arbitrary. Detailed model description is given in the Appendix.

This model was then used to fit the $f(H)$ data for all orientations simultaneously using the least squares method. GNU Octave \cite{octave} software with its standard minimization routines was used for the fitting procedure, the Octave script used for the AFMR frequencies calculations is available at Ref.~\cite{afmr-script}. The resulting best fit is shown in Figure \ref{fig:f(h)} as a solid line. It well describes our experimental data, the best fit parameters are the gaps $\Delta_1=\gamma \sqrt{a_1/\chi_{\perp}}=(13.6\pm 0.1)$~GHz and $\Delta_2=\gamma \sqrt{a_2/\chi_{\perp}}=(5.37\pm 0.05)$~GHz, the gyromagnetic ratio $\gamma=(2.88\pm 0.01)$~GHz/kOe (corresponds to $g=2.06$), the $g$-factor anisotropy parameter $\xi=(0.10\pm 0.02)$ and the angle between the hard axis and the $c^*$ axis $|\phi|=(18\pm 2)^\circ$. We can not determine the direction of rotation from the hard axis toward the $c^*$ axis (clockwise or counterclockwise) from our data. The value of $\Delta_2$ is close to the value of 0.3~K (6.3~GHz) predicted in Ref.~\cite{lederova2017} from the analysis of low-temperature magnetization curves.

Besides of the low-field resonance absorption, we observed a high-field absorption signal at $\vect{H}||a$ (Figure \ref{fig:f(h)}). Taking into account that for $\vect{H}||a$ the field is applied at the angle $\phi$ from the second-easy axis, we fit $f(H)$ dependency for the softening mode as (see Appendix):

\begin{equation}
f=\Delta_{eff} \sqrt{1-(H/H_{sat})^2},
\end{equation}

\noindent here $\Delta_{eff}(\phi)=\sqrt{\Delta_1^2 \sin^2 \phi+\Delta_2^2 \cos^2 \phi}$ ( $\Delta_{eff}=13.1$~GHz for $\vect{H}||a$ experiment).

The best fit value for the saturation field  $H_{sat}=(74\pm 1)$~kOe is slightly larger than the value of 63~kOe determined from the phase diagram of Refs.~\cite{lederova2017,phases-jmmm}. This overestimation of the saturation field is quite natural for the quasi-low-dimensional magnet. The mean-field sublattice model \cite{goorevich} assumes linear magnetization process up to the saturation field with the maximal magnetization $M_0=\chi_\perp H_{sat} $. Magnetization process of low-dimensional magnets is nonlinear with positive curvature at high fields \cite{dejong}, as observed for CUEN as well \cite{lederova2017}, hence real saturation field is less then $ M_0/\chi_\perp$.

\subsection{Microscopic contributions to the anisotropy of the ordered phase and inter-plane ordering pattern}

\begin{figure}
\epsfig{clip=, width=\figwidth,file=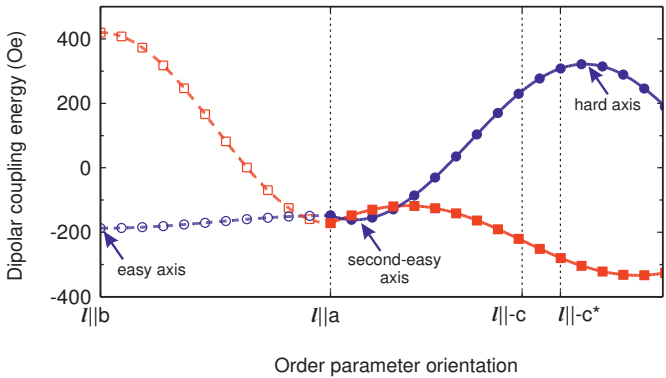}
\caption{Dependency of the dipolar coupling energy per spin (in terms of effective field) on the order parameter orientation for ferromagnetic (blue circles) and antiferromagnetic (red squares) stacking of adjacent $(bc)$ planes. Symbols: computed values, curves: fit of the computed values by $A+B\cos(2(\varphi+\delta))$, the shift $\delta$ is set to zero for the $(ba)$-plane rotation and is used as a fit parameter for the $(ac)$-plane rotation. }
\label{fig:dipolar}
\end{figure}

Experimental identification of the anisotropy axes differs from the predictions of Refs.~\cite{lederova2017,tarasenko2013}. We will discuss below possible microscopic contributions to the anisotropy of the antiferromagnetically ordered state of \cuen{} and will demonstrate that accurate accounting for dipolar coupling explains this controversy and allows to determine inter-plane ordering pattern.

Within the mean-field approach the AFMR gaps are $\Delta_{1,2} = \gamma\sqrt{2H_{A1,A2}H_E}$, where $H_E=H_{sat}/2$ and $H_{A1,A2}$ represent exchange and anisotropy fields \cite{goorevich,note}.  The  effective fields for CUEN are: $H_E = 32$~kOe, $H_{A1} = 0.35$~kOe and $H_{A2} = 0.054$~kOe. While the effective fields are determined with approx. 20\% uncertainty keeping in mind the aforementioned nonlinearity of magnetization process of
two-dimensional CUEN [9], the ratio $H_{A1}/H_{A2} =(\Delta_1/\Delta_2)^2
= 6.41 \pm 0.07$ is determined much more reliably since it does not depend on the exact exchange field value.

Three possible contributions to the anisotropy of the ordered phase can be considered: dipole-dipole interaction, symmetric anisotropic spin-spin coupling (anisotropic exchange interaction) and Dzyaloshinskii-Moria coupling. The symmetric anisotropic interaction was analyzed in Refs.~\cite{lederova2017,tarasenko2013} as the possible source of the anisotropy of static magnetization and of the anisotropic ESR resonance field shift.  Since both $g$-tensor anisotropy and the symmetric anisotropic coupling originate from the same spin-orbit coupling, one can conclude \cite{kubo,lederova2017} that the symmetric anisotropic coupling favors easy plane anisotropy with the main axis of the $g$-tensor ($a$-axis) being the hard axis. The estimate of the anisotropic symmetric coupling constant $\langle G \rangle \simeq 0.02$~K \cite{tarasenko2013} corresponds to the effective anisotropy field  $H_A\simeq\langle G \rangle/(2\mu_B) \simeq 0.15$~kOe (taking into account only four in-plane neighbors).

Estimates of the Ref.~\cite{tarasenko2013} indicated that nearest-neighbor dipolar coupling could provide remarkable contribution to this total anisotropy. To verify this conjecture, firstly, we calculated dipolar contribution to the anisotropy energy of the coupled magnetic layers with intraplane collinear antiferromagnetic order. Strength of the dipolar coupling for nearest neighbors can be estimated as $\mu_B^2/(k_B d_{min}^3) \simeq 4.3$~mK (0.063~kOe in terms of effective field), here $d_{min}=5.27$\text{\normalfont\AA} is the shortest Cu-Cu distance for CUEN. We consider two possible patterns of magnetic $(bc)$ layers stacking:  antiferromagnetic and ferromagnetic stacking  of adjacent magnetic planes (Figure \ref{fig:planes}). Here terms ferro- and antiferromagnetic describe change of the ion magnetization on the  elementary translation $\vect{T}=(\vect{a}+\vect{b})/2$. Dipolar energy was calculated as a function of the order parameter orientation assuming fully saturated magnetization per ion and taking into account known uniaxial $g$-factor anisotropy \cite{tarasenko2013}. Neighbors at the distance up to 100\text{\normalfont\AA} from the given ion were included to dipolar sum. Further increase of the cutoff distance to 150\text{\normalfont\AA} did not change the result. Results are shown in Figure \ref{fig:dipolar}.

Clearly, the ferromagnetic ordering pattern reproduces well the observed anisotropy:  the $b$-axis with the minimum dipolar energy is  the easy axis, the second-easy axis is within 10$^\circ$ from the $a$-axis and the hard
axis is close to the $c$ and $c^*$ axes. The anisotropy energy per spin is $U^{(dip)} = U_0+\mu H^{(dip)}_{A1} l^2_X +\mu H^{(dip)}_{A2} l^2_Y$ (here $\mu$ is magnetization per ion, $X$ and $Y$ are the hard and second-easy axes, $H^{(dip)}_{A1} > H^{(dip)}_{A2} > 0$) with $H^{(dip)}_{A1} = 0.51$~kOe, $H^{(dip)}_{A2} = 0.027$~kOe. The obtained ratio $H^{(dip)}_{A1} /H^{(dip)}_{A2} \approx 19$ is almost threefold higher than the experimentally measured value, therefore the dipolar coupling alone cannot fully describe the observed anisotropy.

Inter-plane exchange couplings calculated from the first principles \cite{firstprinciples} prefer the antiferromagnetic inter-plane stacking: the mean-field  energy for the antiferromagnetic  stacking pattern is 5.6~$\mu$eV per spin less (0.96~kOe in terms of the effective field). However, the dipolar contribution to the anisotropy for the antiferromagnetic stacking pattern completely disagrees with the experiment: the $b$-axis would be the hard axis (Figure \ref{fig:dipolar}).

From the above estimates one can see that the dipolar coupling and inter-plane exchange couplings could be competing in CUEN. In particular, above the spin-flop transition the order parameter is confined to the $(ac)$-plane and the dipolar energy is minimized for the antiferromagnetic inter-plane stacking (Figure \ref{fig:dipolar}). This could result in the rearrangement of the inter-plane stacking pattern above the spin-flop transition

Next, we  consider inter-plane  Dzyaloshinskii-Moriya couplings. ESR linewidth analysis at high temperatures \cite{tarasenko2013,kravchina} demonstrated that only about quarter of the total high-temperature linewidth is due to dipole-dipole couplings (calculated explicitly for CUEN \cite{kravchina}), while the remaining 75\% of the linewidth are more likely due to inter-plane DM coupling. Combination of inversion centers and rotational axes  creates a particular pattern of DM vectors which cancels out effects of DM interaction within the mean-field model.  Inversion symmetry forbids DM coupling on all bonds except for   $J_1$ and $J_2$ bonds (Figure \ref{fig:planes}). $J_1$ bond couples ions within the $(ab)$-plane. Two neighboring $(ab)$-planes (these planes contain ions marked as ``A'', ``B1'', ``B2'' and ``$\alpha$'', ``$\beta 1$'', ``$\beta 2$'', correspondingly, in Figure \ref{fig:planes}) are linked by the inversion, which makes DM vector patterns within these planes exactly opposite: $\vect{D}_{AB1}=-\vect{D}_{\alpha\beta1}$, $\vect{D}_{AB2}=-\vect{D}_{\alpha\beta2}$. Within the same $(ab)$-plane projection of DM vector on the second order axis $b$ alternates: $\vect{D}_{AB1}=(D_X,D_Y,D_Z)$, $\vect{D}_{AB2}=(D_X, D_Y, - D_Z)$, here we use the same XYZ basis  with  $Z$ axis along the $b$-axis and $X$ and $Y$ axes within the $(ac)$-plane. When summed over all bonds within the mean-field model this pattern of DM vectors cancels out exactly and yields no additional anisotropy. $J_2$ bond couples next-nearest layers ($x=0$ and $x=1$ layers in Figure \ref{fig:planes}), due to inversion symmetry DM vectors on bonds originating from ``B1'' and ``$\beta 1$'' ions are opposite, which, again, yields no additional anisotropy.

Now we  can sum up dipolar contribution and contribution of symmetric anisotropic coupling:
\begin{equation}
U=U_0+\mu H_A^{(sym)} l_a^2+\mu H_{A1}^{(dip)} l_X^2+ \mu H_{A2}^{(dip)} l_Y^2,
\end{equation}

\noindent here $l_a$ is the projection of the order parameter on the $a$-axis and $H_A^{(sym)}>0$.
Neglecting  deviation of the second-easy axis $Y$ from the $a$-axis, we found that $H_A^{(sym)}\approx (0.055\pm0.002)$~kOe reproduces experimentally determined ratio of the resulting anisotropy fields. We can estimate anisotropic symmetric coupling constant $ z S^2 G=\mu H_A^{sym}$, here $z=4$ is the number of assumingly equally contributing in-plane neighbors and $S=1/2$. This estimate corresponds to $G\simeq 3.7$~mK, which is in reasonable agreement with conventional estimate of symmetric anisotropic coupling constant as $(\Delta g/g)^2 J$. The total anisotropy fields are then 0.51~kOe and 0.08~kOe, these values are 30\% larger than the values directly estimated from the AFMR experiment ($H_{A1}=0.35$~kOe and $H_{A2}=0.054$~kOe). This scaling can be partially due to the uncertainties in the exchange field definition.

Thus, we can conclude that the dipolar interaction plays the dominant role in the determination of the anisotropy in the ordered phase  of \cuen{} and that the AFMR data can be interpreted in favor of ferromagnetic ordering of the nearest planes.

\subsection{AFMR line splitting}

\begin{table}
\caption{Observed splitting of AFMR line at different field orientations and microwave frequencies measured at the base temperature of 0.45~K}
\label{tab:splitvals}
\begin{ruledtabular}
\begin{tabular}{ccc}
$f$,~GHz& average $H_{res}$,~kOe&$\Delta H$,~kOe\\
\hline
\multicolumn{3}{c}{$\vect{H}||a$}\\
9.6&2.689&$0.014\pm 0.005$\\
11.6&3.824&$0.041\pm 0.005$\\
11.6&36.15&$0.32\pm0.04$\\
14.3&3.157&$0.13\pm 0.015$\\
\hline
\multicolumn{3}{c}{$\vect{H}||b$}\\
4.64&0.75&$0.059\pm0.015$\\
9.6&3.821&$0.019\pm0.005$\\
11.6&4.546&$0.123\pm 0.010$\\
14.3&5.280&$0.037\pm0.007$\\
14.3&5.280&$0.053\pm0.007$\\
17.4&6.368&$0.023\pm0.008$\\
31.6&11.194&$0.021\pm0.005$\\
\hline
\multicolumn{3}{c}{$\vect{H}||c$}\\
14.3&1.560&$0.16\pm0.04$\\
17.4&3.792&$0.063\pm0.015$\\
21.7&5.912&$0.051\pm0.015$\\
\end{tabular}
\end{ruledtabular}
\end{table}

\begin{figure}
\epsfig{clip=, width=\figwidth,file=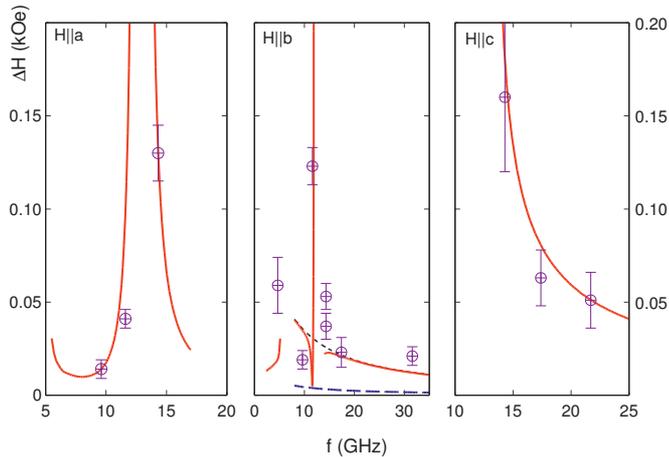}
\caption{Frequency dependency of the observed splitting of AFMR resonance absorption line at the base temperature of 0.45~K (symbols). Solid curves --- model of two decoupled antiferromagnets with slightly different anisotropy parameters and the change of the anisotropy parameters above the spin-flop transition as described in the text. This model includes the $2^\circ$ tilt of the magnetic field at the nominal $\vect{H}||b$ orientation. The dotted curve at the middle panel represents the same model assuming zero tilt at $\vect{H}||b$. The dashed curve corresponds to the model without change of the anisotropy parameters above the spin-flop transition.}
\label{fig:split}
\end{figure}

Observed splitting of the AFMR absorption line below approximately 0.6~K is not expected for the collinear antiferromagnet. Splitting magnitudes are listed in Table \ref{tab:splitvals} and are shown in Figure \ref{fig:split}, the maximal splitting amplitude is about 100-150~Oe. The largest splitting is observed at the frequencies close to the magnon gaps. Whatever the splitting mechanism is, the later observation is quite natural --- the slope of the AFMR frequency-field dependency $df/dH$ approaches zero close to the magnon gaps, thus the small variation of the resonance frequency could result in the substantial change of the resonance field. Several splitting mechanisms can be considered which can be responsible for the aforementioned splitting.

First, we have to consider possibility of the sample twinning. To produce experimentally observed splitting values sample should consist of a mosaic of crystallites rotated by $\sim 10^\circ$. However, the polarized light microscopy of our samples at room temperature does not show any presence of different blocks in the samples. Neither the optical reflectometry at room temperature \cite{optic1,optic2}  revealed  features that could be ascribed to sample twinning. The angular dependencies of the ESR absorption at $T>T_N$ measured at 9~GHz \cite{tarasenko2013} and at 72~GHz (present work) also give no indication of the crystal twinning, while small linewidth of the ESR absorption ($10...20$~Oe at 4.2~K) and $g$-factor anisotropy would make $10^\circ$ rotation of crystallites clearly apparent.  Thus, the sample twining as the source of the  AFMR lines splitting in CUEN can be ruled out decisively.

Another possibility arises from quasi-two-dimensionality of \cuen{}. Two-dimensional Heisenberg antiferromagnet orders only at $T=0$, however, the presence of Ising-type anisotropy results in the ordering at finite temperature. Thus, in the extreme limit of very weak inter-plane coupling, the quasi-two-dimensional antiferromagnet can be considered as a stack of equivalent antiferromagnetically ordered layers weakly coupled by inter-layer Heisenberg exchange interactions. The eigenfrequencies of this system would split similarly to the known textbook problem of coupled oscillators. Such an effect was reported for another quasi-2D antiferromagnet RbFe(MoO$_4$)$_2$ \cite{svistov02,smirnovrbfe}. For the case of field applied along the symmetry axis, split components of resonance absorption correspond  to the in-phase and out-of-phase oscillations of the coupled layers. However, the out-of-phase oscillations of the order parameters correspond to the out-of-phase oscillations of the uniform magnetization of individual layers, which strongly decouples this oscillation mode from the uniform microwave field. At the same time, the experiment (Figure \ref{fig:scans}) shows that the split components have approximately the same integral intensity. This means that corresponding resonance modes are comparably coupled to the microwave field which rules out the Heisenberg inter-layer coupling as the source of the observed splitting.

Finally, we tried a semi-phenomenological model assuming that below $T_N$ two practically decoupled antiferromagnetic systems are formed within \cuen{} lattice. We can recall here the known case of formation of  completely different ordering patterns in the neighboring layers of another quasi-two-dimensional antiferromagnet KFe(MoO$_4$)$_2$ with independent spin dynamics of these layers \cite{kfe-zhel,kfe-jetp}. We will describe these antiferromagnetic subsystems in CUEN phenomenologically by slightly different anisotropy parameters $a_{1,2}$ in Eqn.~(\ref{eqn:UA}):
\begin{equation}
a^{(\pm)}_{1,2}=a_{1,2} (1\pm \delta_{1,2}).
\label{eqn:a-var}
\end{equation}

\noindent The difference of AFMR resonance fields for two antiferromagnets with slightly different anisotropy constants was calculated numerically using best fit values from the Section \ref{sec:f(H)} as  starting parameters. We have found that $\delta_1\neq \delta_2$ is required and the satisfactory description of the data in Figure \ref{fig:split} is achieved for  $\delta_1=0.012$ and $\delta_2=0.005$.

However, this parameters set fails completely to describe splitting at $\vect{H}||b$ orientation above the spin-flop transition (see dashed line in Figure \ref{fig:split}). This can be fixed by assuming that the system-to-system variation of the small anisotropy constant $a_2$ increases almost tenfold in the fields exceeding the spin-flop transition field $H_{SF}$ and can be described now by $\delta_2^{(H>H_{SF})}=0.04$. The increase of the splitting close to the second magnon gap can be accounted by $2^\circ$ tilt of the magnetic field, which is plausible for our experimental setup. The change of the effective anisotropy parameter after the spin-flop transition was earlier reported for the square-lattice antiferromagnet Cu(pz)$_2$(ClO$_4$)$_2$  \cite{povarov} and discussed as a possible effect of dipolar forces \cite{arsenii}. We recall here possibility discussed in previous subsection that inter-plane ordering pattern is rearranged above spin-flop transition due to the difference in dipolar coupling energy, such rearrangement will surely result in the change of anisotropy constants.

From our results we can not reliably conjecture about the origin of the possible formation of two decoupled slightly inequivalent magnetic systems at low temperature. One possible reason is a weak alternation of the crystal structure due to the magnetoelastic coupling leading to the inequivalence of odd and even layers. Verification of this hypothesis requires either a high-resolution structural analysis or a low-temperature NMR experiment to check the number of inequivalent copper ions in the antiferromagnetically ordered \cuen{}.

\subsection{Spin dynamics above the N\'{e}el point}
Phase diagram of \cuen{} was earlier discussed in the context of the field-induced BKT transition \cite{lederova2017,phases-jmmm}. The applied magnetic field  effectively creates an XY anisotropy as the short-range antiferromagnetic correlations develop at $T<\Theta_{CW}$. This field-induced anisotropy competes with thermal effects. In the case of CUEN a crossover from Heisenberg to XY regime takes place below $T_N$ for fields lower than 10 kOe while for higher fields the crossover temperature rises achieving 1.5 K at 20 kOe \cite{phases-jmmm} .

2D magnet with XY anisotropy undergoes a BKT transition, free vortex dynamics above the BKT transition temperature can affect the ESR linewidth leading to characteristic exponential temperature dependency \cite{monika}. Thus, one could expect that ESR linewidth temperature dependency in CUEN would change from critical behavior at low fields (below 10 kOe) to some combination of critical and XY behavior at higher fields while pure XY behavior in sufficiently large temperature interval above $T_N$ can be expected above 40 kOe. However, to distinguish these regimes the ESR linewidths have to be measured very accurately, and even then both scenarios are found to describe experimental data qualitatively well and preferable scenario can be chosen only after quantitative comparison of model parameters with the theoretical predictions \cite{monika}.

To check for these effects we have measured ESR absorption above the N\'{e}el temperature at different microwave frequencies corresponding to resonance field values from 3 kOe to 40 kOe, the latter value is about 2/3 of the saturation field. Accurate determination of ESR linewidth is handicapped in our experimental setup by inhomogeneity of the magnetic field from a compact cryomagnet and distortions of the ESR line shape at high frequencies. We observed ESR linewidth of  20...30 Oe at 1.7K, which broadens up to approx. 200 Oe at $T_N$.
We did not observe strong qualitative difference in temperature evolution of ESR linewidth at different resonance fields which can be interpreted as switching on the additional (vortex) channel of spin relaxation in the field-induced XY regime.

\section{Conclusions}

We have performed detailed low-temperature ESR study of the quasi-2D antiferromagnet \cuen{}. Our results confirm that transition to the magnetically ordered state at 0.9~K is continuous, the ordered phase is a collinear antiferromagnetic state with the easy axis aligned along the crystallographic $b$-axis and the second-easy axis aligned in the $(ac)$-plane at approximately $18^\circ$ from the $a$-axis. Observed anisotropy of the ordered phase can be largely described by dipolar coupling of copper spins, on the base of this analysis the ferromagnetic ordering pattern of the two-dimensional planes is favored.

We have observed additional splitting of the antiferromagnetic resonance absorption spectra which can be interpreted as a coexistence of the two  ordered antiferromagnets with slightly different anisotropy parameters. The microscopic origin of formation of these antiferromagnetic systems is unclear and additional high-resolution structural or NMR experiments are required to get insight on the equivalence or inequivalence of two-dimensional magnetic subsystems of CUEN and on the formed inter-plane ordering patterns.

\acknowledgements
Authors thank Prof. A. I. Smirnov and Prof. L. E. Svistov (Kapitza Institute) for the fruitful discussion and supporting comments.

Authors (V.G. and Yu.K.) acknowledge support of their experimental studies by Russian Science Foundation grant No.17-02-01505 (experiments above 9 GHz) and Russian Foundation for Basic Research grant No. 19-02-00194 (experiments below 9 GHz), data processing and modeling was  supported by Program of fundamental studies of HSE. Work at P. J. \v{S}af\'{a}rik University (R.T. and A.O.) was supported by VEGA grant No. 1/0269/17 of the Scientific grant Agency of the Ministry of Education, Science, Research and Sport of the Slovak Republic. One of the authors (J.Ch.) acknowledge support of his work  by the Path to Exascale project No. CZ.02.1.01/0.0/0.0/16-013/0001791, and by the Ministry of Education, Youth and Sports of Czech Republic project no. LQ1605 from the National Program
of Sustainability II.
\appendix
\section{Equations for antiferromagnetic resonance frequencies}

We  use macroscopic (hydrodynamic) approach of Ref. \cite{andmar}. For the collinear antiferromagnet Lagrangian density is:

\begin{equation}
\label{eqn:lagrangian2}
{\cal L}=\frac{\chi_{\perp}}{2\gamma^2} \left(\dot{\vect{l}}+\gamma \left[\vect{l}\times\vect{H} \right] \right)^2-U_A
\end{equation}

\noindent here  $\vect{l}$ is the collinear antiferromagnetic order parameter, $\gamma$ is the gyromagnetic ratio and $U_A$  is the anisotropy energy. In the case of the monoclinic crystal with the easy axis $Z$ locked to the only second order axis, the anisotropy energy can be written as:

\begin{equation}
U_A=\frac{a_1}{2}l_X^2+\frac{a_2}{2}l_Y^2,
 \end{equation}

 \noindent here $a_1>a_2>0$, $X$ and $Y$ are the directions of the hard and  second-easy axes, respectively.

  The uniaxial $g$-factor anisotropy (as it follows from Ref.  \cite{tarasenko2013}) can be included to anisotropy energy \cite{glazkov-bacusio}:

  \begin{equation}
  U_{A,g}=\xi\chi_\perp\left(\vect{l}\cdot\vect{H}\right)l_a H_a,
  \end{equation}

  \noindent within mean-field model $\xi={\Delta g}/{g}$,  $l_a$ and $H_a$ are the projections of order parameter and magnetic field on the $g$-tensor principal axis (which is $a$-axis for \cuen{}). For $\vect{H}||a$ this term results in the scaling of the gyromagnetic constant to $\gamma_{eff}^2=\gamma^2 \left(1+2\xi \right)$ .

Eigenfrequencies of the order parameter oscillation can be found from the linearized the Euler-Lagrange equations.
This yields two zero-field magnon gaps $\Delta_{1,2}=\gamma \sqrt{{ a_{1,2}}/{\chi_\perp}}$, $\Delta_1>\Delta_2$. At  $\vect{H}||b$ spin-flop transition takes place at $H_{SF}={\Delta_2}/{\gamma}$.

The resonance frequencies for $\vect{H}||b$ are:

\begin{eqnarray}
f_{1,2}^2&=&\gamma^2 H^2+\frac{\Delta_1^2+\Delta_2^2}{2}\pm\nonumber\\
&&\pm\sqrt{2\left(\Delta_1^2+\Delta_2^2\right)\gamma^2 H^2+\frac{\Delta_1^2-\Delta_2^2 }{4}}
\end{eqnarray}

\noindent for $H<H_{SF}$, and

\begin{equation}
f_1^2=\Delta_1^2-\Delta_2^2;~~~
f_2^2=\gamma^2 H^2-\Delta_2^2
\end{equation}

\noindent for $H>H_{SF}$.

For $\vect{H}\perp b$ ($\vect{H}||a, c^*$) secular equation is:

\begin{equation}
\left|
  \begin{array}{cc}
    \gamma_{eff}^2 H_X H_Y & -f^2+\Delta_2^2+\gamma_{eff}^2 H_Y^2 \\
    f^2-\Delta_1^2-\gamma_{eff}^2 H_X^2 & -\gamma_{eff}^2 H_X H_Y \\
  \end{array}
\right|
=0
\end{equation}

\noindent here $\gamma_{eff}=\gamma$ for $\vect{H}||c^*$ and $\gamma_{eff}=\gamma\sqrt{1+2\xi}$ for $\vect{H}||a$.

At high fields one of the AFMR modes asymptotically approaches Larmor frequency $\gamma H$ and the frequency of the other AFMR mode remains small (it does not exceeds the larger gap $\Delta_1$) and softens at the saturation field \cite{goorevich}.  While the hydrodynamical approach is not directly applicable at high fields, it can be shown that at high fields the angular dependency of the low-frequency AFMR mode resonance frequency is a universal function for a given antiferromagnet  \cite{far}. Thus, we can calculate the asymptotic frequency of the low-frequency mode within the low-field hydrodynamic approach for the field $\vect{H}\perp Z$:
\begin{equation}\label{eqn:gap-far}
\Delta_{eff}(\phi)=\sqrt{\Delta_1^2 \sin^2 \phi+\Delta_2^2 \cos^2 \phi},
\end{equation}

\noindent  here angle $\phi$ is counted from the hard axis. Combining Eqn.~(\ref{eqn:gap-far}) with the predictions of the sublattices model  \cite{goorevich} one obtains resonance frequency of the low-frequency AFMR mode:
\begin{equation}
f=\Delta_{eff} \sqrt{1-(H/H_{sat})^2}.
\end{equation}


\begin{thebibliography}{10}
\bibitem{note} In the case of uniaxial anisotropy, the anisotropy energy per magnetic ion $U_A =K l^2_z$ corresponds to the anisotropy field $H_A = K/\mu$, here $\mu$ is the magnetization of the ion.
\end{thebibliography}
\end{document}